# Self-Organized Networks and Lattice Effects in High Temperature Superconductors II: Fermi Arc Anomalies

J. C. Phillips

Dept. of Physics and Astronomy, Rutgers University, Piscataway, N. J., 08854-8019

## Abstract

The self-organized dopant percolative filamentary model, entirely orbital in character (no fictive spins), explains the evolution with doping of Fermi arcs observed by ARPES, including the previously unexplained abrupt transitions in quasiparticle strength observed near optimal doping in cuprate high temperature superconductors. Similarly abrupt transitions are also observed in time-resolved picosecond relaxation spectroscopy at 1.5 eV, and these are explained as well, using no new assumptions and no adjustable parameters.

**1. Introduction**

High temperature cuprate superconductivity may be the most complex phenomenon known in inorganic materials. It has been the subject of more than 65,000 papers, and a large number of theoretical models have attempted to explain the many counter-intuitive phenomena observed. Many of the theoretical papers contain elaborate and ingenious mathematical models whose relation to experiment is vague. In this paper topological methods, based on the author's earlier theories of the glassy behavior of dopants in the cuprates [1], are used to discuss several anomalies in detail, with emphasis on the key role played by the connectivity of the internal dopant structure. The theory emphasizes qualitative trends, as experience has shown that that the complexity of these materials may well preclude quantitative treatments of the kind that worked so well for simpler superconductors, such as $MgB_2$.

This paper is part of a longer paper that has been broken into parts to make it more digestible. (Certainly it is much more accessible than the 65,000 papers that have been written on the cuprates!) I have retained the Section, Figure and Reference numbering of the longer paper to avoid translational errors. The first part of this paper appeared as cond-mat/0611089.

**5. Broad Implications of Phase Diagrams**

It is customary, in many theoretical papers, to begin an elaborate mathematical discussion by showing a token phase diagram borrowed from an experimental paper, such as the diagram shown in Fig. 3 [36]. The diagram shows several phases: an antiferromagnetic AF phase near dopant concentration $x = 0$, an intermediate phase that is superconductive and exhibits non-Fermi liquid normal-state transport properties, a pseudogap phase, and a Fermi liquid nonsuperconductive phase at large $x$. There is, however, much more in this phase diagram than merely a rich variety of phases.

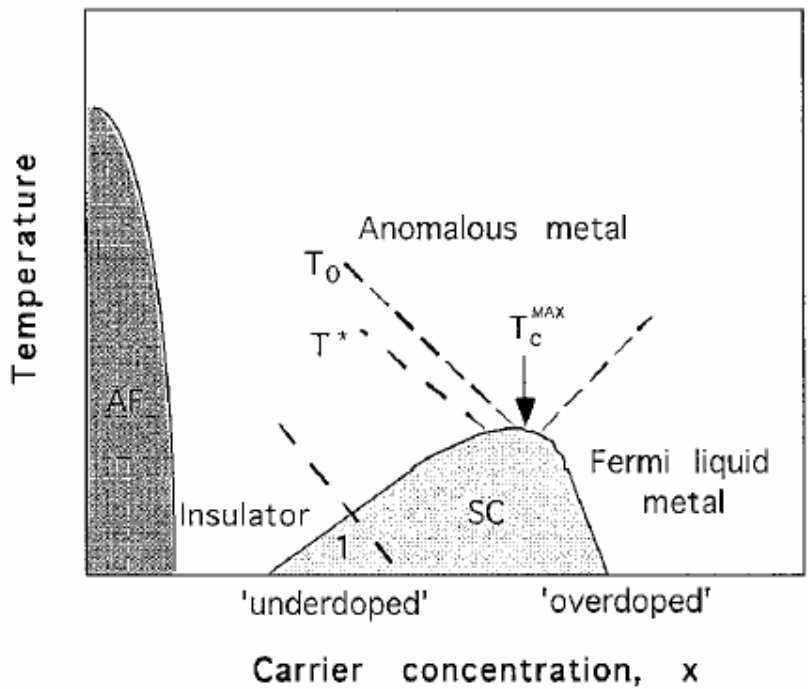

Fig. 3. *A sketch of the "standard" phase diagram for HTSC, including lines for the pseudogap defined in terms of magnetic susceptibility and electrical resistivity [38].*

First, there is the question of the relevance of the AF phase to the superconductive phase, as the two are separated by the pseudogap phase. There is no doubt that NiO is a strongly AF material over a wide range of alloying with other oxides, but in the cuprates the AF phase disappears at small values of x (~ 0.03 in $La_{2-x}Sr_xCuO_4$ (LSCO), see below). Why does this happen? As we have seen in Fig. 2(a), the cuprates are marginally stable elastically. The AF instabilities associated with half-filling of the Cu d $(x^2 - y^2)$-Op(x,y) conduction band compete with, and are easily replaced by, charge-density waves based on 2x1, 2x2, etc. reconstructions, and this will occur at very small values of x because of the very large numbers of degrees of freedom in the 3 nm glassy nanodomains observed by STM [5]. In other words, the pseudogaps in the regions that form the maze through which dopant-based filaments percolate are almost surely CDW pseudogaps, not AF pseudogaps.

This is a very important observation, because theories are appearing that seem to resemble the present theory, yet are fundamentally different from it. For example, a recent 2006 theory [37] uses as its subtitle the title of my uncited 1989 book [18], and it discusses topological aspects of its model, but the topology is completely different. The model assumes that the host lattice is AF, and that the dopants are not interstitial, but are placed on lattice sites. This translationally invariant all-lattice spin model, which treats phonons largely as merely a thermal nuisance, is similar to spin-glass models of network glasses, which beg the question of self-consistent network connectivity and tell us little or nothing about the physics of real off-lattice glasses like window glass [1,22]. Electric dipoles have generated filamentary paths in successful off-lattice simulations of the intermediate phase (also not cited in [37]). The simulations exhibit quite generally the combined effects of space-filling and competing long-and short-range forces in the formation of the intermediate phase [38]. In the AF model these space-filling off-lattice electric dipoles have been replaced by spins on a lattice; of course, lattice models cannot explain the formation of the intermediate phase! Moreover, magnetic spins have pseudovector symmetry, while electric dipoles are vectors. We shall see in latter sections that the mysterious magnetic precursive effects observed in HTSC are in fact closely correlated to lattice strains indicative of strong off-lattice interactions of electric dipoles. The latter first order into loops and then pass through a mean-field topological ordering transition near the onset of superconductivity. On behalf of the AF theory [6,37], it should be said that, unlike string theory (where the physics is "not even wrong"), its physics is "completely wrong". Alternatively, the mathematics of both string theory and [37] are "interesting" [6], but not "amusing". Many other popular "toy" models of HTSC, such as t-J and Hubbard models, are also mathematically "interesting", but not "amusing".

There is much more to be learned from the standard cuprate phase diagram by comparing it to the phase diagrams of other materials that undergo superconductor-insulator transitions, notably thin strongly disordered W-Si films [38], and the $(Ba,K)(Pb,Bi)O_3$ family [39]. In both cases $T_c$ increases to its largest value as the transition is approached with decreasing doping. The dopants are largely disordered, and the EMA is valid. The increase occurs because near the transition the screening of the attractive electron-phonon interaction breaks down, and at the transition some changes occur in the internal structure that open an insulating gap. There are indications of some internal ordering of the K dopants (incipient breakdown of the EMA) in the $(Ba,K)BiO_3$ case [39] with increasing $T_c$.

Ando *et al.* obtained very sophisticated features of the cuprate phase diagram by studying the T dependence of the planar resistivity in the normal state [40] (Fig. 4). Plots of the ab planar resistivity $\rho_{ab}(x,T)$ of $La_{2-x}Sr_xCuO_4$ exhibit *two* Ando lines (I,II) $d^2\rho(x)/dT^2 = 0$. Line I is diagonal and defines $T_p(x)$, while line II defines a critical crossover very nearly fixed vertically at optimal doping, x = 0.15. Line I has a corner near x = 0.06, where the metal-insulator transition occurs; it is possible that this corner is also connected with a crossover from an AF pseudogap to a CDW pseudogap (see above). The *rectilinear* (not curvilinear) nature of both Ando lines is strong evidence for the percolative character of the (super)conductive paths formed by the internal filamentary network, as such rectilinearity is characteristic of percolative phenomena in strongly disordered (glassy) systems [41]. The nature of the critical crossover (line II) at optimal doping is revealed by the experiments (using samples prepared in Ando's lab) that are the subject of the next Section.

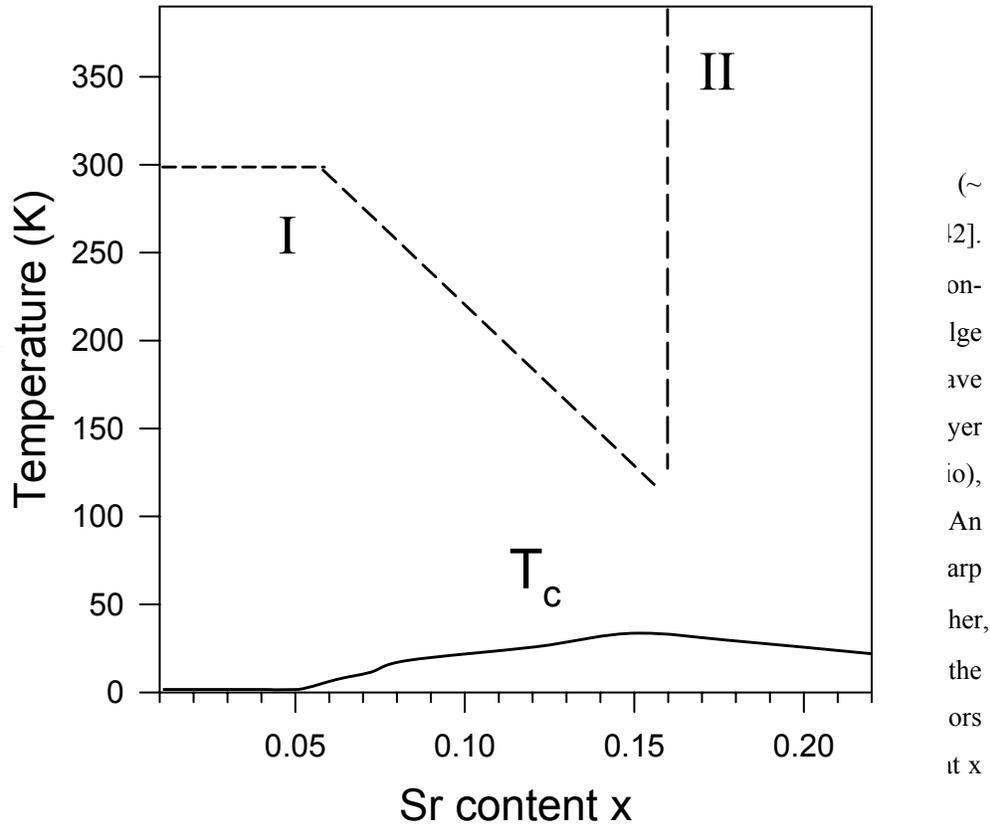

Fig. 4. The planar resistivity $\rho(x,T)$ phase diagram of $La_{2-x}Sr_xCuO_4$ (LSCO), after Fig. 2(c) of [40]. Two lines zeroes of $d^2\rho(x)/dT^2$ are marked I and II, as discussed in the text.

There are other interesting aspects of the cuprate phase diagram, notably the bulge (~ 0.1%) in the planar lattice constants of $La_{2-x}Sr_xCuO_4$ in the intermediate phase [42].

One might have thought that the superconductive condensation due to attractive electron-phonon interactions would have caused a lattice contraction in the xy plane, so this bulge could be surprising. However, we saw above that the strong attractive s wave interactions occur at the interlayer dopants, and these cause contraction of the interlayer spacing. Since the unit cell volume remains nearly constant (negative Poisson's ratio), the interlayer contraction causes a bulge in the planar lattice constants [12]. An interesting feature of the bulge is that at optimal doping, $x = 0.15$, it reaches a sharp minimum at $T_0 = 190$ K [43]. The significance of $T_0$ will be discussed later. Further, there is also a positive bulge (~0.1%) in the temperature dependence below 150K of the mean square *relative* displacements determined by the correlated Debye-Waller factors $\sigma^2$ of the Cu-O bonds by EXAFS at $x = 0.1$, which is measurably weaker (< 0.03%) at $x = 0.2$ [44]. The meaning of this difference between under- and over-doped networks is discussed later in Sec.6.

**6. Emerging Fermi Arcs and Crossovers at Optimal Doping**

HTSC characteristically exhibit three phases, an insulating phase (dopant concentration x in the range $0 < x < x_1$), an intermediate superconductive phase with anomalous normal-state transport up to temperatures ~ 700K ($x_1 < x < x_2$), and a non-superconductive metallic or Fermi-liquid phase ($x > x_2$) [1]. The superconductive transition temperature $T_c(x)$ reaches a maximum near the center of the intermediate phase at the optimal doping $x = x_0$.

ARPES experiments [45] on $La_{2-x}Sr_xCuO_4$ (LSCO, $x_1 = 0.06$, $x_2 = 0.22$, $x_0 = 0.16$) showed metallic behavior in the Fermi arc even at $x = 0.03$ (insulating phase) beginning in the nodal $(\pi,\pi)$ gap direction, and spreading smoothly over a wider angle as x increased. Moreover, the intensity of the metallic signal exhibited an *abrupt* (essentially step function, hence unavoidably nonanalytic) qualitative change at $x = x_0$, with the largest value shifting from the nodal $(\pi,\pi)$ gap direction to the antinodal $(\pi,0)$ gap direction for $x > x_0$. The spreading behavior is absolutely incompatible with conventional analytic models based on Fermi liquid theory, which would predict that in the metallic phase the Fermi arc would be present at all angles. Moreover, addition of dopants would shift the Fermi arc radially (rigid band model), but no such shifts have been observed. It is evident that conventional analytic models cannot explain the ARPES data (especially the *abrupt* qualitative change), and that a new kind of model, based on novel mathematical methods, is needed.

In this section our topological approach shows that abrupt qualitative changes occur at optimal doping because this is the composition at which the network consisting of separated filaments, each with its own phase-coherent currents in the normal state, fill the space between the pseudogap barriers without overlapping and losing their separate phase coherence. Space-filling is a topological concept inaccessible to analytic models. In lattice models it is introduced trivially, which conceals its self-consistent character.

By now it is well known that Fermi liquid (mean field) theory does not provide an adequate description of the electronic spectra of cuprate high temperature superconductors measured by ARPES. There are many difficulties: the Fermi arc does not shift with doping, but instead even in the insulating phase with $x = 0.03 < x_1 = 0.06$, states begin to appear at $E_F$ in the nodal directions [45], and with increasing x an occupied arc $\mathbf{k}(E_F)$ centered on this direction grows in width and intensity $Z(x,\alpha,E_F)$, where the angle $\alpha = \pi/4$ (0) for the gap nodal (antinodal) directions. The growth is not completely smooth (there are jumps in $Z(x,\pi/4,E_F)$ when x crosses $x_1$ and between x = 0.10 and 0.15, Fig. 5(a)), but the most striking feature of this Fermi arc is a very large jump in $Z(x,0,E_F)$ when x crosses $x_0$ (goes from underdoped to overdoped, and $T_c$ goes through its maximum; see Fig. 5(b)). Because it is able to explain both the origin of the intermediate phase and the origin of d-wave gap anisotropy, topological constraint theory is ideally suited to discussing the abrupt transition in $Z(x,0,E_F)$ for $x = x_0$ observed in ARPES data [2]. For the reader's convenience the main features of these data are sketched in Fig. 5. (The ARPES data are quite complex and should be viewed in full color [45].)

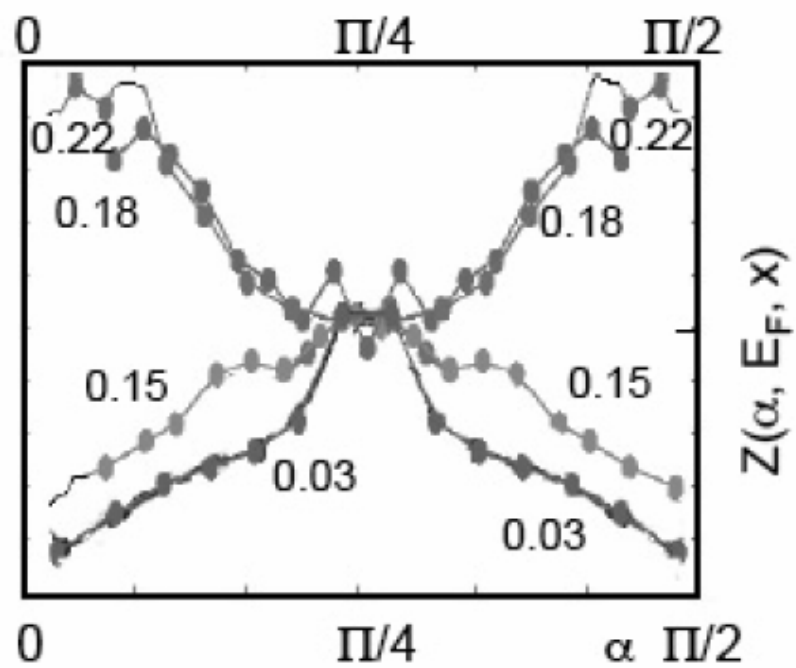

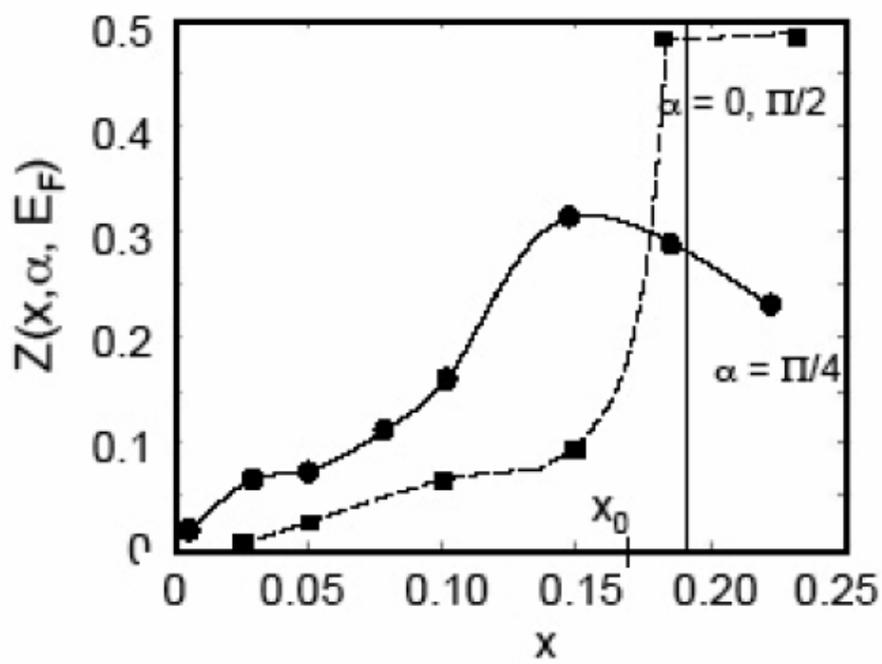

*Fig. 5. (a) A sketch of Z(x,α,E$_F$) (in arbitrary units, different for each curve) for La$_{2-x}$Sr$_x$CuO$_4$ (LSCO) [45] for x = 0.03, 0.15 (almost optimal doping), 0.18 and 0.22 (both overdoped); [45] also gives data for x = 0.05, 0.07, and 0.10. The central feature of these data, explained here, is the abrupt jump between x = 0.15 and x = 0.18 that is especially large near α = 0, π/2, but is small near α = π/4. This difference is illustrated in Fig. 5(b).*

The hierarchical topological explanation for these unprecedented phenomena is that the highly mobile dopants self-organize to form conductive (electronically coherent) segments even in the insulating phase when annealed at high temperatures. By forming such conductive segments the dopants lower the sample free energy by better screening internal electric fields in this strongly ionic material. The anisotropy of the energy gap tells us that the antinodal (nodal) directions α = 0 (π/4) are the directions of strongest (weakest) electron-phonon interactions. Thus the nodally oriented segments have the lowest scattering and the highest conductivity, and the segments that are formed at small x = 0.03 are therefore oriented predominantly in the α = π/4 direction. With increasing x the segments begin to percolate at x = x$_1$ to form a metallic network. Because of the presence of a high density of Jahn-Teller distorted insulating islands (large pseudogaps [5]) the percolative paths become sinuous, and to avoid filamentary crossings, the occupied Fermi surface arc spreads to larger values of │α -π/4│ with increasing x. Note that these filamentary states are states bound or pinned to the dopants outside the metallic planes. The host momentum spectrum remains that of the metallic patches in the metallic planes, which is why the Fermi arcs do not move in **k** space as x increases and dopants are added to insulating layers. Note that the filamentary paths are three-dimensional, and are not confined to the metallic planes; when projected onto these planes, the filamentary states lie in the energy gaps of the nanodomain walls. Without these walls the filamentary bound states could not be formed.

Turning now to the most striking feature of the ARPES data, the very large jump in $Z(x,0,E_F)$ when x crosses $x_0$ (see Fig. 5(b)), the authors [45] explain this anomaly as arising from a flat energy band arising from states near $(\pi,0)$. However, it is not easy to see how such an energy band could be derived from a periodic potential. It will generate a very large peak in $N(E)$, the density of quasi-particle states, at $E = E_F$. Normally such a peak will split into two peaks, above and below $E_F$ (Jahn-Teller effect). Moreover, it would have to be narrow on an energy scale of order 0.01 eV. Within our model such states are easily pinned in the nanodomain wall energy gap, but without such a gap it seems to be impossible to generate such a flat band.

The flat band of states pinned to $E_F$ at $x = x_0$ must also consist of states localized in **k** space near $(\pi,0)$. In a band model these extra states would presumably be associated with a saddle point in **k** space located at $(\pi,0)$, but such a model cannot account for the abruptness of its appearance. In our model it is easy to see how this happens. As x increases towards optimal doping $x = x_0$, the weak scattering directions near $\alpha = \pi/4$ on the Fermi arc are filled by dopants with local filamentary tangents with $\alpha$ orientations, and finally only the strong scattering directions near $\alpha = 0$ are left; these are filled last. Beyond $x = x_0$, additional dopants must occupy sites close enough to other dopants that strong scattering occurs at these dopant pairs. This strong scattering locally destroys filamentary coherence, giving rise to Fermi liquid (incoherent) patches. Because the most highly conductive directions are the $\alpha = \pi/4$ directions, the internal fields will be best screened by orienting the overdoped doping pairs along $\alpha = 0$ directions. The structure factor $S_2(\mathbf{k})$ of these closely spaced pairs then reflects the "flatness" (in **k** space) of the states responsible for the abrupt jump in $Z(x,0,E_F)$ when x crosses $x_0$ (see Fig. 5(b)). The dopant pairs are larger than the individual (atomic scale) dopants that form the underdoped filaments responsible for the nodal arc; the natural length scale for these pairs is the nanodomain diameter, of order 10 unit cells [5]. The structure factor of the overdoped pairs is thus ~ 10 times more localized in **k** space than the underdoped nodal structure. Also one should note that in the present model there is a sum rule: $\int d\alpha\, Z(x,\alpha,E_F) \sim x$. (To implement this rule one should include the dependence of state broadening $\Gamma(x, \alpha)$ in estimating Z from the data.) Thus the abrupt increase of $Z(x,0,E_F)$ when x crosses $x_0$ takes place at the expense of $Z(x,\alpha,E_F)$ for values of $\alpha$ not close to 0.

While the foregoing discussion justifies the usage of the term "flat bands" to explain the abruptness of the appearance of the strong intensity of incoherent antinodal states beyond $x = x_0$, it is important to realize that these states are not one-electron band states in the usual sense of rigid or fixed energy bands.  Dopants close enough together to form Fermi liquid droplets develop quadrupole moments with the long axis oriented along the Cartesian axes; these clusters are many-electron systems, characterized by strongly anisotropic one-electron scattering.  As soon as such droplets begin to form, $T_c$ begins to decrease, as filamentary coherence is disrupted by Fermi liquid incoherence.  The (in)coherence arises from the local topology of the dopant clusters, and it is not a continuum property, even though it may be probed by ARPES photons with wave lengths large compared to the dopant spacing and nanodomain dimensions.

**7. Comparison with other theories**

Unlike analytical models, topological models are not adorned by elaborate algebra, but they also contain *no adjustable parameters*.  Their great strength lies in their ability to identify and even predict the essential qualitative features of complex problems; here these problems arise from the need to identify the essential features of many-body dopant configurations relevant to a given experiment.  The Fermi arcs identified by ARPES have attracted great theoretical interest, previously discussed using analytical models; it is therefore useful to compare these analytical models with the present topological model, particularly with respect to qualitative features.  Although by now it is abundantly obvious that electron-phonon interactions cause high temperature superconductivity, and that gas- or liquid-like quasiparticle models cannot account even qualitatively for the dependence of the phonon kink on x and isotopic mass [46,47], almost all the mean field models that claim to explain the Fermi arc analytically in underdoped cuprates invoke spin (not phonon) degrees of freedom, usually in the context of multiply parameterized t-J or Hubbard models with generous helpings of uncontrolled renormalization of two-dimensional energy gaps by three-dimensional interactions (another adjustable parameter) [48]. Strong claims are made for these models [49,50]: they contain the "essential" physics of the cuprates, the momentum dependence of their quasiparticle states is "essentially" that observed by ARPES, etc. Similarly strong claims have been made for Fermi liquid models with strong spin relaxation channels [51].

A striking feature of all the spin-dependent mean field (Fermi liquid) models is that they purport to explain optical data (which reflect almost entirely electric dipole transitions, with oscillator strengths determined by orbital coherence) in terms of both spin and orbital degrees of freedom, yet they never mention the awkward fact that the atomic spin-orbit coupling $\lambda(\mathbf{L}\cdot\mathbf{S})$ parameters $\lambda$ for O 2p and Cu 3d states are ~ 0.01 eV, which is negligibly small compared to the orbital band widths W ~ 1 eV. In these "strong coupling" Hubbard mean field models the spin and orbital degrees of freedom are coupled as artifacts of projection procedures driven by the Coulomb repulsion parameter U, assumed to be ~ 10W. These "central cell" (short range) projection procedures involve many uncontrolled approximations that are probably incompatible with coherent intercellular (long range) orbital motion, and the gradual appearance of the Fermi arc may be merely an artifact of the projection procedures used. (Note that in the parallel classical problem of mechanical elasticity phase changes from floppy to rigid, in the mean field approximation *there is no intermediate phase* [52]. It seems likely that were the quantum models to be treated exactly in mean field, the intermediate phase would disappear there as well; it appears as an artifact derived from smoothing out a first-order mean field phase transition.) In particular, there appears to be no way that these *smooth* projection procedures can generate the *abrupt* transition in $Z(x,0,E_F)$ when x crosses $x_0$ at optimal doping observed in ARPES (Fig. 5(b)). Of course, no justification has ever been given for abandoning the ideas of self-consistent one-electron calculations for these uncontrolled approximations, which to my knowledge have never been confirmed experimentally for any carefully studied simple material. These self-consistent one-electron calculations are the basis for all successful predictions of optical and photoemission spectra of solids [53,54], and they have successfully predicted matrix element effects in cuprate photoemission spectra [55]. If the reader scans this manuscript for the phrase "space filling", she will find many applications of the space-filling concept that are simply not accessible by analytic methods.

**8. Picosecond reflectance relaxation**

The emphasis in the present self-organized (adaptive, not random) percolation model on formation of dopant filaments apparently requires a "Maxwell demon" to identify such dopants, especially as optical experiments involve wave lengths much longer than the nanodomain dimensions identifiable only by large area, high-resolution scanning tunnel microscopy (STM) studies [5]. However, time-resolved relaxation studies have proved to be a powerful tool in studying network self-organization in both molecular and electronic glasses [1,56], and elegant pump-probe reflectance relaxation studies [57] at 1.5 eV have revealed picosecond anomalies in $La_{2-x}Sr_xCuO_4$ (LSCO, $x_1 = 0.06$, $x_2 = 0.22$, $x_0 = 0.16$) that closely parallel the ARPES anomalies, with an abrupt change in relaxation time at optimal doping.

It seems surprising that a simple reflectance measurement, which averages over all **k** values involved in interband optical transitions, can yield an anomaly very similar to that obtained by the much more refined and precise ARPES **k**-resolved technique; this is possible because there are some very subtle aspects to pump-probe time-resolved relaxation studies in self-organized glassy systems [56]. These are illustrated in Fig. 6(a) and 6(b). The electron-hole pair created by photon absorption relaxes rapidly (femtosecond time scale) to a metastable excitonic state which then relaxes much more slowly, probably non-radiatively (phonon emission), on a picosecond time scale. The experimental relaxation data [57], reproduced for the reader's convenience in Fig. 7(a),(b), refer to this second regime. There are not one, but two, surprises in these data. First, $\Delta R(x)/R$ smoothly reverses sign at $x = 0.16 = x_0$, and second, the relaxation time $\tau$ abruptly drops from 20 ps in the underdoped regime $x < 0.16 = x_0$ to ~ 2 ps for $x \geq 0.16$. As emphasized in [57], this is strongly reminiscent of the abrupt crossover in $Z(x,\pi/4,E_F)$ - $Z(x,0,E_F)$ in Fig. 5(b). At the same time, it is hard to understand how this abrupt drop can be associated with a peak in a quasiparticle density of states $N(E)$ pinned to $E_F$ at $x =$ $x_0$. (In a rigid band quasiparticle model $\tau$ would reach a minimum at $x = x_0$ and then recover for both $x < x_0$ and $x > x_0$.) The self-organized (not random) filamentary percolation model, with its strong local field corrections, readily explains both of these results (surprising and *basically inexplicable in a mean field quasiparticle context*).

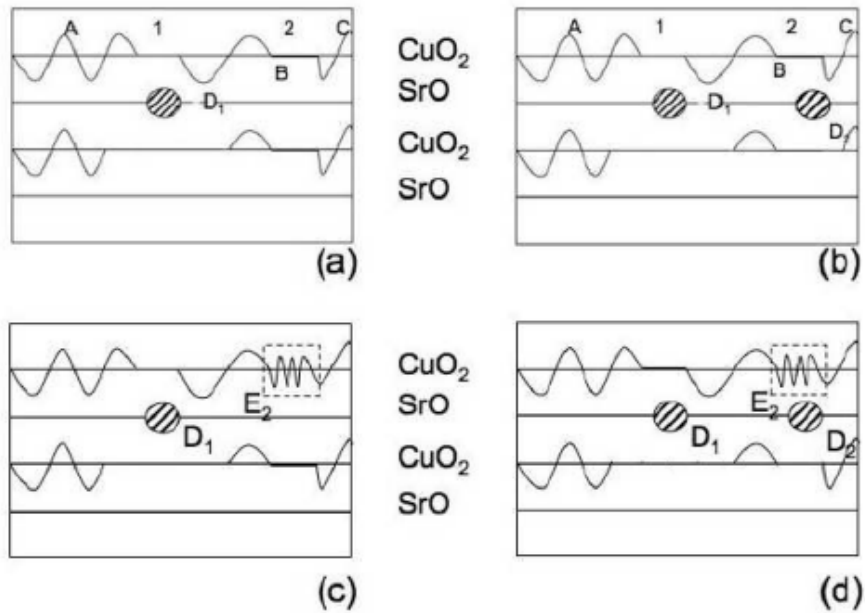

*Fig. 6. The percolative model of Fig. 1 is redrawn here to illustrate the effects on path coherence due to crossover from underdoped to overdoped. The metallic regions of the CuO₂ planes are indicated by wavy curves; these regions are separated by semiconductive nanodomain walls, indicated by straight lines. In (a) we have two such walls, marked 1,2. In wall 1 there is a dopant in the SrO layered labeled $D_1$. A coherent filamentary current path can go from A to B in the CuO₂ plane by utilizing $D_1$ as a bridge to bypass wall 1. The filament ends at B. In (b) a second dopant $D_2$ has been added to wall 2, so that now the filament continues from A through B to C. In (c) $D_2$ has been replace by a surrogate excitonic bridge $E_2$; this structure explains the positive value of*

*ΔR/R observed in underdoped samples in picosecond relaxation experiments [54]. Finally in (d) we have the overdoped case, where the exciton overlaps a dopant, locally destroying the coherence of the filament, and giving rise to a negative value of ΔR/R.*

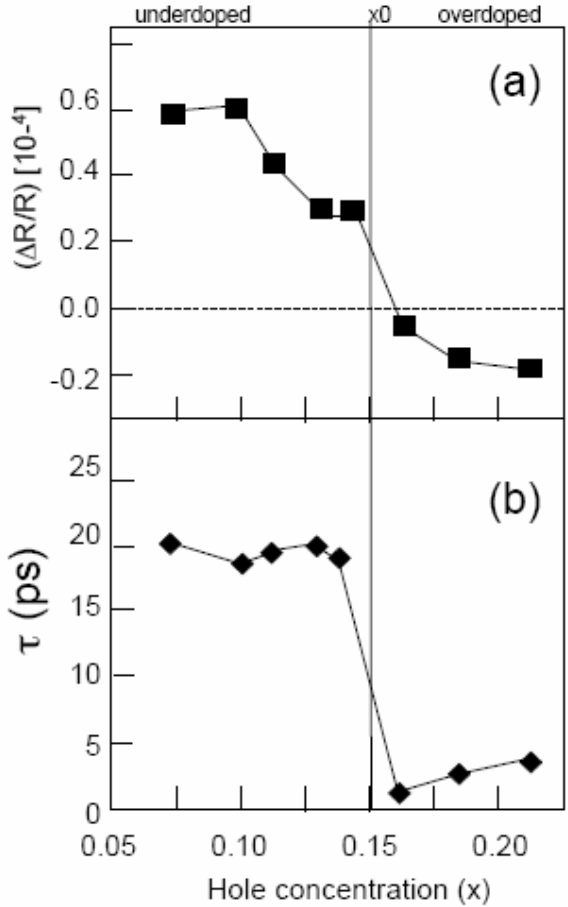

*Fig. 7. For the reader's convenience the relaxation data from [54] are reproduced here: (a) ΔR(x)/R changes smoothly with x, and reverses sign at $x = x_0$; (b) τ(x) changes abruptly (essentially a step function) at $x = x_0$. The ps time scale of τ(x) implies that phonons play an essential part in the relaxation. Because the relaxation is essentially dielectric relaxation, spins are irrelevant.*

Because the pump and probe both have energy 1.5 eV, this experiment detects changes that occur because of addition of a metastable exciton.to the filamentary network. The exciton is metastable because it has been added to a semiconductive nanodomain wall (excitons added to the metallic nanodomains decay on the femtosecond time scale), and it decays by "leaking out" (tunneling) to the metallic regions. The positions occupied by the metastable exciton are not all equivalent, even though the probed single-particle exciton energy remains fixed at the pump energy 1.5 eV. The exciton is highly polarizable, and the free energy of the filamentary network depends on the position of the exciton. For $x < x_0$, the exciton diffuses to positions that minimize the system's free energy by increasing the conductivity of the filamentary network. Thus the exciton functions as a surrogate dopant, increasing the length and connectivity of coherent filaments, and hence increasing the reflectivity, $\Delta R/R > 0$. However, for $x > x_0$, the filaments have already filled the available phase space (optimized glass networks are space-filling and "incompressible"), and when excitons are added, they function again as dopants, but this time as excess dopants, disrupting network segments and producing incoherent Fermi-liquid like regions.

There are two reasons why $\Delta R(x)/R$ changes smoothly with x. First, the high conductivity sites, with $\alpha = \pi/4$ filamentary tangents, are occupied first, and the lower conductivity sites (increasing $|\alpha - \pi/4|$) are then filled with increasing x. Second, note that for $x < x_0$, the exciton must diffuse from its initial region to find the best filamentary site, and that this is an easier task for strong underdoping, when the filaments are nearly oriented in $\alpha = \pi/4$ channels, than near optimal doping, when the filaments have become sinuous, and the entire network must reconfigure itself to improve its coherent conductivity. For $x > x_0$, the disruption is minimized for surrogate insertion near $\alpha = 0$ oriented filamentary tangents, where the local conductivity is less than in $\alpha = \pi/4$ channels. That is why the magnitude of $\Delta R/R$ is smaller for strong overdoping than for strong underdoping. One can add that a semiclassical simulation of coherent filamentary percolation could be carried out in the context of directed percolation, with angular weighting factors. Such a simulation would be considerably more complex than the mechanical simulations of [52], and it lies outside the framework of this paper.

The situation for $\tau(x)$ is different. For $x < x_0$, decay of the exciton must occur through a Franck-Condon configurational barrier (the exciton abandons its surrogate coherent filamentary function, and decays into an incoherent state), which leads to a large $\tau$. For $x > x_0$, the exciton decays from an incoherent state in the semiconductive wall to another incoherent state in the adjacent metallic nanodomain. Because both states are incoherent, the Franck-Condon configurational barrier almost disappears, and the decay takes place on a ps (one phonon) time scale (ten times faster than for $x < x_0$).

The measured abrupt drop in the average value of $\tau$ by a factor of 10 at $x = x_0$ can be estimated as follows. According to the discussion at the end of Section 2, in the metallic region $x > x_0$, the energy level spacing in a nanodomain is of order $W/N$, where $W \sim 1$ eV is the valence band width, and $N$ is the number of unit cells in the nanodomain. In the filamentary region $x < x_0$, the intrafilamentary energy level spacing in a nanodomain is of order $W/N^{1/3}$, so it is much smaller. All other things being equal, Fermi's golden rule for transition rates says that the ratio of the relaxation times will scale as $N^{2/3}$. This gives $N \sim 30$, which is consistent with the observed nanodomain dimensions of 3 nm [5].

A related effect, superconductivity-induced spectral weight shift $\Delta SW_{sc}(x)$ below 1.25 eV in BSCCO, also is a skewed parabola that tracks the parabolic $T_c(x)$ [58]. This behavior is exactly what is predicted by the three-phase filamentary model: from underdoped to optimally doped $\Delta SW_{sc}(x)$ increases as the number of filaments increases to its space-filling limit. In the overdoped samples $\Delta SW_{sc}(x)$ decreases rapidly, mirroring 1.5 eV pulse-probe reflectivity and relaxation time drops [57], because of interfilamentary scattering in Fermi liquid patches.

This paper benefited greatly from the presentations and discussions at the workshop **http://cnls.lanl.gov/Conferences/latticeeffects/** sponsored by Los Alamos National Labs.